\documentclass[conference]{IEEEtran}
\usepackage{amsmath}
\usepackage{amsfonts}
\usepackage{amssymb}
\usepackage{hyperref}
\usepackage{fixltx2e}
\usepackage{listings}
\usepackage{graphicx}
\usepackage{pdflscape}
\title{Jif: Language-based Information-flow Security in Java}
\author{\IEEEauthorblockN{Kyle Pullicino}
\IEEEauthorblockA{School of Informatics\\
University of Edinburgh\\
Edinburgh, United Kingdom \\
Email: s1461410@sms.ed.ac.uk}}
\date{November 7, 2014}

\begin{document}
\lstset{language=Java,basicstyle=\scriptsize\ttfamily,tabsize=1}
\maketitle
\begin{abstract}
In this report, we examine Jif, a Java extension which augments the language with features related to security. Jif adds support for security labels to Java's type system such that the developer can specify confidentiality and integrity policies to the various variables used in their program. We list the main features of Jif and discuss the information flow problem that Jif helps to solve. We see how the information flow problem occurs in real-world systems by looking at two examples: Civitas, a ballot/voting system where voters do not necessarily trust voting agents, and SIF, a web application container implemented using Jif. Finally, we implement a small program that simulates information flow in a booking system containing sensitive data and discuss the usefulness of Jif based on this program.
\end{abstract}
\begin{IEEEkeywords}
security,
information flow,
java,
jif,
confidentiality,
integrity
\end{IEEEkeywords}
\section{Introduction}
\emph{Jif} is a Java extension which augments the language with features related to security. Primarily, it helps developers enforce information flow security constraints at the code level by using specific Jif annotations and constructs. Jif is available on \url{http://www.cs.cornell.edu/jif/} and in this report we review the extension. In this report we aim to,

\begin{itemize}
\item give a detailed description of the main features offered by Jif and the problems that they solve,
\item implement a practical example using these features and
\item critique the Jif extension with regards to its suitability as a solution for security related application concerns.
\end{itemize} 

We first give an overview of Jif and its features including how it augments Java's type system so that developers can explicitly state confidentiality and integrity policies for the data in their programs. We also look at how Jif can be used for two real-world cases, namely a voting/ballot system and a general web framework.

We also present a small program implemented using Jif which will also be used for critiquing Jif towards the end of the report.

\section{Related Work}

Jif is one of multiple projects that attempt to solve the information flow security problem. FlowFox, for example, is a web browser implemented to take information flow into consideration and protect the user's sensitive data \cite{de_groef2014}. A type system for a modified version of JavaScript has been proposed by Hedin and Sabelfeld to introduce information flow security in web scripts \cite{hedin2012}. Both these proposals are relevant because they encounter the same problems which will be described later in this report and which apply similarly to Java.

The idea of a security label lattice which is subsequently used by Jif comes from Denning \cite{denning1976}. Denning proved that a lattice can be used to verify the information flow security of a program. Denning \& Denning also point out that this verification can be done statically by a compiler \cite{denning1977} which is the primary motivation for developing information flow type systems in languages.

Sabelfeld and Myers discuss information flow security as it pertains to programming languages in their paper and also point out the more subtle problems related to language-based information flow \cite{sabelfeld2003}. These include the fact that modern languages are becoming increasingly more expressive and more concurrent.

\section{Background: The Jif extension}

In this section, we introduce Jif and its main features. All the information here is from the Jif reference manual available online \cite{jif}. We will focus on the basic features that Jif provides and how they can be used to solve common security issues when developing software applications.

\subsection{Information flow}

The information flow problem is a common problem in software development concerning the leaking of sensitive information to unauthorised people or entities. This typically occurs not because of a malicious intent by the software developer but due to bugs and errors introduced in the system during development stage. A web application might inadvertently keep sensitive information in memory longer than necessary which is later leaked out as a response to a user request. This is clearly a very dangerous issue and tools like Jif are meant to help tackle and reduce the likelihood of something like this occurring.

\subsection{Principals, security policies \& labels}

Jif is concerned about information flow between so-called \emph{principals}. Principals are the entities involved with the system. They can represent a single human user, for example, or even a whole group of users. Jif defines a relation between principals called the \emph{acts for} relation. For instance, if a given principal, Alice, can act for another principal, Bob, (written $Alice \succeq Bob$) then Bob delegates all of his authority to Alice. Jif also defines a \emph{top} principal, $\top$, which acts for all principals ($\forall p \colon \textbf{Principals} \cdot \top \succeq p$) and a \emph{bottom} principal, $\bot$, which allows all principals to act for it ($\forall p \colon \textbf{Principals} \cdot p \succeq \bot$).

Jif extends the Java type system so that types can be declared along with a \emph{security label}. This security label describes a pair (ie. \emph{2-tuple}) made up of a confidentiality policy and an integrity policy. Each policy is owned by a \emph{principal}. A confidentiality policy specifies which principals the owner allows to read a particular data variable. Likewise, an integrity policy specifies which principals the owner allows to modify a piece of data. For example, the confidentiality policy, $Alice \rightarrow Bob$, means that the data to which this policy applies is owned by Alice and can be read by Alice herself, Bob and any other principal that can act for any of them. Likewise, $Charles \leftarrow Bob$ is the integrity policy, owned by Charles, which allows Charles, Bob and anyone who can act for them to modify the associated data.

By combining a confidentiality policy together with an integrity policy, we get a security label of the form $\{c;d\}$ where $c$ is a confidentiality policy and $d$ is an integrity policy. A label is added to a variable declaration to specify the security policies that govern access to the contained data. So, for example, we can declare an integer variable which everyone can read but only Alice can write to as follows,
\begin{center}
\texttt{int\{Alice->\_; Alice<-*\} secret;}
\end{center}

\subsection{Method labels}

During execution, Jif maintains a \emph{program counter} which keeps track of the most restrictive security policy in effect. If the program attempts to access data with a less restrictive policy than the one in the program counter, the compiler will generate an error warning the developer that information is being leaked at that particular point in the program.

By using a begin-label on a method, we instruct Jif to check that the program counter has the appropriate security policy before execution starts inside the method. It also enforces that write attempts that happen inside the method conform to the active integrity policy.

An example of a method declaration with labels, taken from the Jif reference manual, is shown below,

\begin{center}
\begin{verbatim}
public void setElementAt{L}(Object{L} o, 
     int{L} i) { }
\end{verbatim}
\end{center}

\texttt{L} here is a previously defined label parameter which works similarly to generics. The begin-label of the method \texttt{setElementAt} is given in braces just after the method name.

\subsection{Authorities}

A principal may also give authority to a method to act for it using the \texttt{authority} construct. An example from the Jif reference manual is given below,

{
\small
\begin{verbatim}
class Game authority(referee) {
    void start() where authority(referee) {
        // this entire method body has the 
        // authority of referee
        ...
    }
}
\end{verbatim}
}

In this snippet, the class declares that its methods can act for a principal called \texttt{referee} in its signature. Then, each one of its methods may add that it acts with the authority of \texttt{referee} by adding the clause \texttt{authority(referee)} in its signature. Now, each time the system queries the active security policy in the program counter, it will do so with the authority of the \texttt{referee} principle.

\subsection{Other features}

Jif supports other advanced features which we will not cover in this report as we will not use them in the demonstration in section \ref{sec:demo}. Notably, Jif supports polymorphism with parametrized classes (which works similarly to generics), dynamic labels which store label information at runtime and extensible principals allowing developers to define their own principal types.

\section{Jif in action}

We now look at two example applications of Jif. Each paper referenced here has been published by Jif's creators to promote Jif. We review their arguments for applying Jif in two particular situations with commonly occurring real-world equivalents.

\subsection{Civitas: a voting system}

\emph{Civitas} is a voting system built using Jif \cite{civitas2007}. The authors state that in an election, both integrity of the whole balloting process and confidentiality of the individual votes are necessary. They say that, traditionally, each one of these can be solved at the expense of the other. So, for example, integrity can be achieved during an election if everyone publicly states their own vote; this method, though, forgoes confidentiality because everyone's vote becomes public knowledge.

Civitas makes use of a log service shared across four different types of voting agents. Voting agents are the entities running and organising the election process separate from the electorate. A log service allows data insertion signed using a secure digital signature (prevents forging additional messages). The bulletin board, used by the agents to tally votes, and the individual ballot boxes, used by voters to submit their vote, are all instances of the aforementioned log service.

Using Jif, security policies can be enforced on all the various components of the arrangement mentioned above to ensure that no unintended information flow occurs in the system.

In the system mentioned above, a registration teller agent issues credentials which a voter must use when submitting their vote. When the credentials are created in the system, the registration teller then labels the credentials with a confidentiality policy of $RT \rightarrow voter$. This means that the owner of the policy (the RT or registration teller principal) also allows the voter principal to read the credentials.

Another example given by the authors is that of the integrity policy $TT \leftarrow Sup$ applied to ballot boxes. This label means that the tabulation teller (TT) will consider that the ballot box's integrity has been compromised if someone else other than the supervisor (Sup) or the tabulation teller themselves has affected the value of the ballot box.

Using a modified version of Jif called Jif\textsubscript{E}, the system also declares multiple \emph{declassification} and \emph{erasure policies} which state the conditions that must hold before a confidentiality policy can be relaxed or made more restrictive. In the Civitas system, each registration teller must store a private credential share that will be requested by the voter as described above. By using an erasure policy, the authors declare that the registration teller must erase the private share component once it is handed to the voter. In this way, we have a provably correct piece of code which is clearly destroying the sensitive data: there is no way that the private credential component will be leaked to unintended entities because the erasure policy mandates its destruction.

\subsection{SIF: Servlet Information Flow framework}

\emph{SIF} is a framework built using Jif, on top of the Java Servlet Framework, used for creating servlet-based web applications \cite{chong2007}. Chong et al. discuss how web applications suffer from multiple threats due to their nature. Specifically, web applications must frequently communicate with potentially unknown clients. These clients are not necessarily benign and may take advantage of vulnerabilities to retrieve sensitive data which they are not authorised to access.

A well known information flow attack is \emph{SQL injection} where a user sends SQL commands to the web application which are sent straight to the underlying RDBMS server without any sanitisation. The RDBMS server will execute the unsanitized command in full and possibly send back information which should not have been accessible to the user.

Jif, instead, protects against more general attacks of a similar nature. A user might know the URL of a particular page on the web application containing sensitive data. Due to human error, the page might have been left accessible without the need for authentication. Using Jif, the sensitive data itself (as opposed to the web page displaying that data) can be labelled and Jif will ensure that such data never leaves the web application as a response unless the authenticated principal is correct. This is somewhat similar, albeit on a smaller scale, to the example implementation given later in section \ref{sec:sensitive}.

SIF makes heavy use of dynamic principals since web applications usually add new users (through some sort of registration process) during their lifecycle. Dynamic principals are a recent addition to Jif; they allow Jif's type checker to reason about principals which are not yet fully known in advance during compilation.

The authors give an example of a web application, implemented in SIF, providing a calendar service for its users. Users create events which they own and, using different security policies, can share their events with other users and attendees. Using Jif's integrity policy labels, SIF allows users to express whether they want other users to modify events or just see events that they shared. At a lower level, the system is checking whether the authenticated user can act for an event's creator or one of the attendees (using a special Jif operator, \texttt{actsfor}).

By labelling the appropriate data in this way, the compiler will statically check that an event object's information never reaches a user who is not the owner or one of the attendees. A person who is reading the code finds a proof of security in the labels that are specified with each object's declaration.

\section{A Jif Example}
\label{sec:demo}

In this part of the report, we present a small program written in Jif to demonstrate how security labels can be used to restrict unauthorised information flow. The full program listing is found in Appendix \ref{apx:demo}.

\subsection{Defining sensitive data}
\label{sec:sensitive}

For this program, we simulate a small part of a tour booking system. A class \texttt{Booking} represents a booking object created by either one of two users of the system, namely Alice and Bob. Each booking object has a sensitive field of type \texttt{String} called \texttt{cardNumber} which is the 16-digit credit card number used by the customer when paying for the booking. It is extremely important for a system of this sort that the card number is never shown to an unauthorised user. We define the \texttt{Booking} class as well as the \texttt{cardNumber} string as shown in listing \ref{lst:booking}.

\begin{figure}[h]
\begin{lstlisting}[caption={Booking class with sensitive card number.},label={lst:booking}]
public class Booking[principal Owner, 
	principal Operator] authority(Owner) {
    private final String{Owner->*} cardNumber;
    
    // ... Getter methods, etc.
}
\end{lstlisting}
\end{figure}

The class is defined with respect to two principal parameters, the \texttt{Owner} who created the booking and an \texttt{Operator} user who is managing the bookings using the system.

By adding a label \texttt{\{Owner->*\}} to the \texttt{cardNumber} variable, Jif will make sure that \texttt{cardNumber} is never accessed by any principal which is not the the owner of the booking. In fact, the getter function of this variable (\texttt{getFullCardNumber()}) must also have its begin-label set to \texttt{\{Owner->*\}}: the compiler will check that the program counter has the appropriate security level before any calls to the getter method can go through.

\subsection{Declassifying confidential information}

Information sometimes needs to flow from a highly restricted and secure domain to a less restricted domain as part of the business specifications of the system itself. In Jif, this is realised as moving data from a variable with a particular label to another variable with a less restrictive label.

\begin{figure*}
\centering
\includegraphics[width=0.85\textwidth]{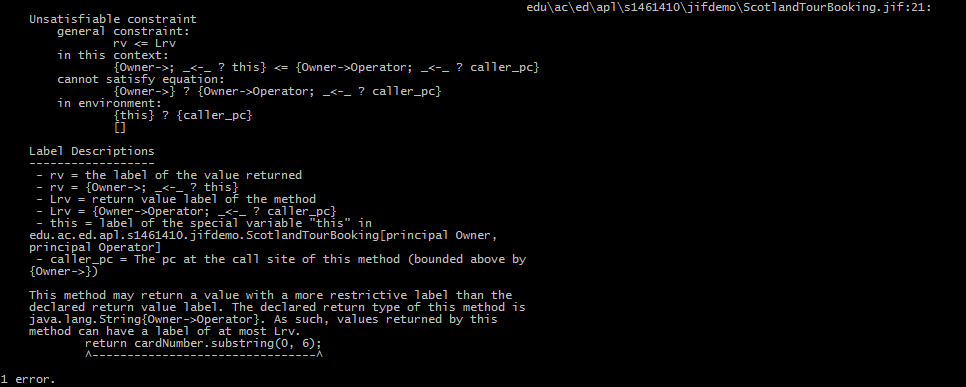}
\caption{Compiler error when card number is not declassified first.} \label{fig:error}
\end{figure*}

The Jif compiler would normally not allow such an operation and will give an error. An example of this is shown in figure \ref{fig:error} which is an error given when attempting to compile the method shown in listing \ref{lst:error}. \texttt{getFirstSix()} is a method which will return the first six digits of the card number used to pay for the booking. The specifications for this program state that an operator user can look at the first six digits of a card number without compromising the full card number.

\begin{figure}
\begin{lstlisting}[language=Java,caption={Incorrect flow of information. Jif will not compile this method.},label={lst:error}]
public String{Owner->Operator} getFirstSix{Owner->*}() {
   try {
        return cardNumber.substring(0, 6);
   } catch (Exception e) {
        return "N/A";
   }
}
\end{lstlisting}
\end{figure}

Jif will not compile the method because its return type's label is \texttt{\{Owner->Operator\}} which is less restrictive than the card number's label \texttt{\{Owner->*\}}. For this to work, we must explicitly declassify the sensitive data as shown in listing \ref{lst:declassify}.

\begin{figure}
\begin{lstlisting}[language=Java,caption={Using \texttt{declassify} to relax the restrictions on the card number variable.},label={lst:declassify}]
public String{Owner->Operator} getFirstSix{Owner->*}() : 
	{Owner->Operator} where authority(Owner) {
   try {
	String{Owner->Operator} result = "";
	result = declassify(cardNumber, {Owner->*} to 
		{Owner->Operator});
	return result.substring(0, 6);
   } catch (Exception e) {
        return "N/A";
   }
}
\end{lstlisting}
\end{figure}

The \texttt{declassify} keyword forces Jif to downgrade the security policy to a less restrictive one. Notice how the programmer must specify which policy label they want to downgrade from as well as the policy label they are downgrading to. This is Jif's way of ensuring that the developer is fully aware whenever information flow occurs from a domain with high security to a less restrictive one.

To prevent the developer from declassifying any arbitrary data, declassifying a policy requires the authority of the owner of that policy. For a confidentiality policy (ie. $A \rightarrow B$), the owner of the policy is the principal that appears on the left hand side. So, for the policy label that is attached to the card number, \texttt{\{Owner->*\}}, the method \texttt{getFirstSix()} requires the authority of the \texttt{Owner} principal as specified in the method's signature.

Furthermore, anyone reading a program in Jif can clearly look for these boundaries where information is allowed to escape (by looking for \texttt{declassify} keywords) and check that the program is indeed doing what it is meant to be doing according to some business specification.

\subsection{Protection against unauthorised access}

We now look at the main application class of our system which performs operations on the booking class we defined in the previous sections. This method demonstrates how Jif protects information from inadvertently escaping its security domain. The \texttt{execute()} method is given in listing \ref{lst:execute}.

\begin{figure}
\begin{lstlisting}[caption={The \texttt{execute()} method in the \texttt{Application} class.},label={lst:execute}]
public void execute{Alice->Chuck meet Bob->Chuck 
	meet Chuck->*}() 
		where authority(Alice, Bob, Chuck) {
	Booking[Alice, Chuck]{Alice->Chuck} booking1 = 
		new Booking[Alice, Chuck]("4444333322221111");
	Booking[Bob, Chuck]{Bob->Chuck} booking2 = 
		new Booking[Bob, Chuck]("4444333322221111");
        
	String{Alice->*} aliceNotebook = 
		booking1.getFullCardNumber();
	// The compiler would issue an error for the next line
	// String{Bob->*} bobNotebook = 
	//	booking1.getFullCardNumber();
        
	String{Chuck->*;Alice->Chuck} operatorNotebook 
		= booking1.getFirstSix();
}
\end{lstlisting}
\end{figure}

The program first creates two bookings, one for Alice and one for Bob. Chuck is the system's operator. Each booking takes, as principal parameters, the owner and operator users in square brackets. We then simulate a request for data by filling in a ``notebook'' for each of the users in the system.

Alice writes the full card number she used when placing a booking in her notebook, with label \texttt{\{Alice->*\}}. Since this is the same label used for the Alice's card number in her booking object, Jif allows the operation and the statement compiles.

The commented line contains a statement where Bob tries to copy Alice's card number from her booking object into his own notebook. The label on Bob's notebook, \texttt{\{Bob->*\}} is less restrictive than the label on Alice's card number. In this case, Jif would issue an error at compile time similar to the error in figure \ref{fig:error}. Jif stops us from making the critical mistake of inadvertently passing Alice's card number to Bob.

Finally, Chuck writes Alice's first six digits from her card number in his notebook. Now, since this is the same information that is coming from Alice's card number, the label on Chuck's notebook must be a conjunction of Chuck's confidentiality policy and Alice's confidentiality policy on her card number's first six digits\footnote{This is an ideal example of where we could have considered using an integrity policy instead of combining confidentiality polices. We would keep the operator's notebook readable only by Chuck but its contents can be affected by Alice: $\{Chuck\rightarrow \top;Chuck\leftarrow Alice\}$.}

\section{Reflections on Jif}

Our reflection and opinions regarding Jif are based on our experience while working on the example given in the previous section. By using Jif, we believe that protecting crucial sensitive data is much easier provided that the labels are used correctly.

Jif's labelling system suffers from the same problems apparent in any other type system. Namely, the type system will work as long as the developer uses it correctly. As a type system, such as Jif's, becomes more complex, the likelihood of making an error\footnote{``Error'' here in the sense that the user does not fully specify the correct types and takes short-cuts.} becomes greater.

While working on the example presented in this report, we encountered multiple issues with regards to selecting the correct labels for the program to compile. Admittedly, this is mostly due to our inexperience of working with Jif and trial-and-error did eventually get the job done. Jif's lack of widespread adoption and community using the tool is also partly to blame. A more time-constrained developer would have taken the short-cut of simply removing any label constraints which were proving to be a problem until the program compiles: the resultant system would supposedly be resilient to incorrect information flow but in reality, none of Jif's security features come into play.

In spite of this, we still believe that Jif is a suitable tool with a lot of promise. It has a good span of features which are all documented. Jif's creators should now focus more on expanding the toolset available for Jif (a Jif file editor with code suggestion is sorely needed, for instance). Also, additional documentation should be provided for troubleshooting common problems as a new programmer using Jif will likely encounter multiple problems when using it for the first few times.

\section{Conclusion}

In this report, we discussed the information flow problem, a security concern for systems which deal with sensitive data. We gave an overview of the \emph{Jif} extension for Java which is a tool to help alleviate this security concern. We listed its main features and how they are expressed in Java code.

\emph{Civitas} and \emph{SIF} were given as examples of systems built using Jif where we see the information flow problem clearly. Civitas, for example, has multiple entities which have to cooperate together but are mutually distrusting of each other. Jif therefore enforces security constraints on the data which the aforementioned entities have to work with. Likewise, SIF implements security constraints for web applications which are repeatedly targeted in information flow attacks where an adversary attempts to extract unauthorised information which they normally should not have access to.

Finally, we developed a small example using Jif where we protected sensitive data from accidentally leaking outside its security domain. Using the program we also gave constructive criticism for Jif as a whole. In essence, we have come to the conclusion that Jif is indeed a suitable tool for solving the information flow problem.

\bibliography{report}{}
\bibliographystyle{plain}

\newpage
\renewcommand{\thepage}{\Roman{page}}% Roman numerals for page counter
\setcounter{page}{1}
\appendix
\section{Demo Program Listing}
\label{apx:demo}

\subsection{Booking.jif}
\begin{lstlisting}[language=Java]
package edu.ac.ed.apl.s1461410.jifdemo;

public class Booking[principal Owner, 
    principal Operator] authority(Owner) {

    private final String{Owner->*} cardNumber;
    
    public Booking(String{Owner->*} number) {
        this.cardNumber = number;
    }

    public String{Owner->*} getFullCardNumber{Owner->*}() {
        return cardNumber;
    }
    
    public String{Owner->Operator} getFirstSix{Owner->*}() 
            : {Owner->Operator} where authority(Owner) {
        try {
            String{Owner->Operator} result = "";
            result = declassify(cardNumber, 
                {Owner->*} to {Owner->Operator});
            return result.substring(0, 6);
        } catch (Exception e) {
            return "N/A";
        }
    }
    
    public String{Owner->Operator} getLastFour{Owner->*}() 
            : {Owner->Operator} where authority(Owner) {
        try {
            String{Owner->Operator} result = "";
            result = declassify(cardNumber, 
                {Owner->*} to {Owner->Operator});
            return result.substring(12, 16);
        } catch (Exception e) {
            return "N/A";
        }
    }
    
    public String{Owner->Operator} getHashedNumber{Owner->*}() 
            : {Owner->Operator} {
        return getFirstSix() + "******" + getLastFour();
    }
    
}
\end{lstlisting}

\subsection{Application.jif}
\begin{lstlisting}[language=Java]
package edu.ac.ed.apl.s1461410.jifdemo;

public class Application authority(Alice, Bob, Chuck) {
    
    public void execute{Alice->Chuck meet Bob->Chuck meet Chuck->*}() 
            where authority(Alice, Bob, Chuck) {
        Booking[Alice, Chuck]{Alice->Chuck} booking1 = 
            new Booking[Alice, Chuck]
            ("4444333322221111");
        Booking[Bob, Chuck]{Bob->Chuck} booking2 = 
            new Booking[Bob, Chuck](
            "4444333322221111");
        
        String{Alice->*} aliceNotebook = 
            booking1.getFullCardNumber();
        // The compiler would issue an error for the line below
        // String{Bob->*} bobNotebook = 
        //  booking1.getFullCardNumber();
        
        String{Chuck->*;Alice->Chuck} operatorNotebook = 
            booking1.getFirstSix();
    }
    
    public static void main{Alice->Chuck meet Bob->Chuck meet Chuck->*}
            (String[] args) {
        new Application().execute();
    }
    
}
\end{lstlisting}

\end{document}